\documentclass[onecolumn,secnumarabic,amssymb, nobibnotes, aps, prd]{revtex4}
\usepackage{amsmath} \usepackage{amssymb}
\usepackage{graphicx} 
\usepackage{epstopdf}
\usepackage{amsmath}
\usepackage{amsmath,amssymb,amsthm,amsfonts,mathrsfs,bm,verbatim}
\usepackage{graphicx,subfigure}

\newcommand{\bea}{\begin{eqnarray}}
\newcommand{\eea}{\end{eqnarray}}

\begin{document}

\title{Thermodynamic geometric analysis of conventional black holes under f(R) gravity }%

\author{Wen-Xiang Chen$^{1,a}$}
\affiliation{Department of Astronomy, School of Physics and Materials Science, GuangZhou University, Guangzhou 510006, China}
\author{Yao-Guang Zheng$^{2,a}$}
\affiliation{Department of Astronomy, School of Physics and Materials Science, GuangZhou University, Guangzhou 510006, China}
\author{Jing-Yi Zhang$^{1}$}
\email{zhangjy@gzhu.edu.cn}
\affiliation{Department of Astronomy, School of Physics and Materials Science, GuangZhou University, Guangzhou 510006, China}
\thanks[a]{These authors have contributed equally to this work.}

\begin{abstract}
In this article, we consider a conventional black hole under f(R) gravity. Due to the small fluctuations around the equilibrium state, we have summarized the expression of the modified thermodynamic entropy of this black hole. At the same time, we also studied the geometric thermodynamics (GTD) of black holes, and studied the adaptability of the curvature scalar of the geometric thermodynamics method to the phase transition point of the black hole. In addition, we also studied the influence of the modified parameters on the thermodynamic behavior of black holes.

 \text { KEYwords: } f(R) \text { gravity; thermodynamic geometry; black hole. }

\end{abstract}

\maketitle

\section{Introduction}
The study of the thermodynamic properties of black holes has always been an important and very attractive research topic in gravitational physics.\cite{1,2,3,4,5,6} In particular, according to the AdS/CFT correspondence, the study of black hole thermodynamics under the background of AdS space-time has received great attention. In this space, Hawking and Page discovered that there is a stable large Schwarzschild black hole to the hot gas space phase transition, the so-called Hawking-Page phase transition. According to the AdS/CFT correspondence, this phase transition is interpreted as the confinement/uncontainment phase transition of the strong coupling gauge theory. What's more interesting is that under the background of a charged or rotating AdS black hole, there will be a phase transition between a small black hole and a large black hole, which enriches people's understanding of black hole thermodynamics. In recent years, the research hotspot of black hole thermodynamics under the background of AdS space-time is to regard the cosmological constant as the thermodynamic pressure of the black hole thermodynamic system, and the total energy of the black hole as the enthalpy of the system. After this interpretation of the cosmological constants, the thermodynamic laws of the AdS black hole system completely correspond to the laws of the ordinary thermodynamic system, and more and more studies have shown that the black hole system contains a rich phase structure. In this article, we will also use this interpretation of cosmological constants to study the thermodynamic properties of fine black holes and their corresponding phase transition properties.

The study of black hole thermodynamic phase transition from the perspective of thermodynamic geometry is an important method in the study of phase transition\cite{4,5,6,7,8}. Therefore, we will construct three different thermodynamic geometries for this fine black hole in the parameter space based on the Hessian matrix, namely Weinhold geometry, Ruppeiner geometry, and free energy geometry. For these three different geometries, we will calculate their geometric standard curvatures separately. Furthermore, according to the specific form of the scalar curvature, the divergence behavior is discussed in depth, and the phase transition point and critical point of the black hole are compared to reveal the relationship between the phase transition point of the black hole and the singular curvature behavior.

Ruppeiner geometry is used in this article. Ruppeiner geometry\cite{4,5,6,7,8} can reveal some characteristics of statistical mechanics to some extent, so we are of course very interested in applying it to black hole thermodynamics. Because most of the black hole information is unknown to us, the statistical mechanics model behind its thermodynamics, except for a small part of the black hole in the superstring or M theory. Ferrara et al. introduced the method of thermodynamic geometry for the first time to study the thermodynamics of black holes. It is found that the Weinhold metric is proportional to the metric of the extreme black hole in the supersymmetric mode space. The temperature of this black hole is zero, and the Ruppeiner metric diverges in the theory of extreme thermodynamics of fluctuations. Some authors have found that when the Ruppeiner metric is considered as a function of mass and charge, the metric of the RN black hole is always flat, and the curvature scalar is equal to zero. When the Ruppeiner metric is considered as a function of entropy on mass and angular momentum, the metric of the Kerr black hole is always curved, and the curvature scalar diverges in the extreme case of the Kerr black hole. In other words, the Ruppeiner metric of the RN black hole is different from the Ruppeiner metric of the Kerr black hole.

In the context of $f(R)$ modified gravitational theory, we study the Kerr-Newman black hole solution. We study the non-zero constant scalar curvature solution and discuss the metric tensor that satisfies the modified field equation. We study the thermodynamics of BHs and their local and global stability without cosmological constants. We analyzed these characteristics in various $f(R)$ models. We commented on the main differences in general relativity and demonstrated the rich thermodynamic phenomenology that characterizes the framework.

In this article, we obtained the modified thermodynamic entropy and studied the thermodynamic quantities and thermodynamic geometric methods of the black hole in $f(R)$ gravitation. We found that the Hawking temperature is a decreasing function of the radius of the horizon. In order to evaluate the correction of static black hole entropy in $f(R)$ gravity caused by thermal fluctuations, we use the expressions of Hawking temperature and uncorrected specific heat.

\section{Thermodynamic parameters of Schwarzschild black holes in f(R) theory}

 In this section, we summarize the expression of the black hole correction thermodynamic entropy caused by small fluctuations near the equilibrium. For this, let us first define the density of states with a fixed energy as(with natural unit, $G=\hbar=c=1$)\cite{9,10}
 \begin{equation}
\rho(E)=\frac{1}{2 \pi i} \int_{c-i \infty}^{c+i \infty} e^{\mathcal{S}(\beta)} d \beta
\end{equation}
 The exact entropy $\mathcal{S}(\beta)=\log Z(\beta)+\beta E$ clearly depends on the temperature $T\left(=\beta^{-1}\right)$. So, this (precise entropy) is not just its equilibrium value. The exact entropy corresponds to the sum of the entropy of the thermodynamic system subsystems. The thermodynamic system is small enough to be considered in equilibrium. In order to study the form of exact entropy, we solve the complex integral by considering the method of steepest descent around the saddle point $\beta_{0}\left(=T_{H}^{-1}\right)$ so that $ \left.\frac{\partial \mathcal{S}(\beta)}{\partial \beta}\right|_{\beta=\beta_{0}}=0 .$ Now, the Taylor expansion of the exact entropy is performed around the saddle point$\ beta=\beta_{0}$ causes
 \begin{equation}
\mathcal{S}(\beta)=\mathcal{S}_{0}+\frac{1}{2}\left(\beta-\beta_{0}\right)^{2}\left(\frac{\partial^{2} \mathcal{S}(\beta)}{\partial \beta^{2}}\right)_{\beta=\beta_{0}}+(\text { higher order terms })
\end{equation}
 and
 \begin{equation}
\rho(E)=\frac{e^{\mathcal{S}_{0}}}{2 \pi i} \int_{c-i \infty}^{c+i \infty} \exp \left[\frac{1}{2}\left(\beta-\beta_{0}\right)^{2}\left(\frac{\partial^{2} \mathcal{S}(\beta)}{\partial \beta^{2}}\right)_{\beta=\beta_{0}}\right] d \beta
\end{equation}
we get that
\begin{equation}
\rho(E)=\frac{e^{\mathcal{S}_{0}}}{\sqrt{2 \pi\left(\frac{\partial^{2} \mathcal{S}(\beta)}{\partial \beta^{2}}\right)_{\beta=\beta_{0}}}}
\end{equation}
where $c=\beta_{0}$ and $\left.\frac{\partial^{2} \mathcal{S}(\beta)}{\partial \beta^{2}}\right|_{\beta=\beta_{0}}>0$ are chosen.

Using the Wald relationship between the Noether charge of the differential homeomorphism and the entropy of a general spacetime with bifurcation surfaces, we introduce a method to obtain the effective family of higher-order derivatives from the entropy of black holes. We take the entropy as the starting point and analyze the derivation process of the action functional\cite{17}.

We briefly review the static $f(R)$ black hole solution and its thermodynamics. The generic form of the action (in unit $G=c=\hbar=\kappa=1$ ) is given by
\begin{equation}
I=\frac{1}{2} \int d^{4} x \sqrt{-g} f(R)+S_{\text {mat }}
\end{equation}
among them, $S_{\text {mat }}$ refers to the movement of gravity and the material part of the expression is considered to be\cite{17}
\begin{equation}
f(R)=R-q R^{\beta+1} \frac{\alpha \beta+\alpha+\epsilon}{\beta+1}+q \epsilon R^{\beta+1} \ln \left(\frac{a_{0}^{\beta} R^{\beta}}{c}\right)
\end{equation}
Where $0 \leq \epsilon \leq \frac{e}{4}\left(1+\frac{4}{e} \alpha\right), q=4 a_{0}^{\beta} / c( \beta+1), \alpha \geq 0, \beta \geq 0$ and $a_{0}=l_{p}^{2}, a$ and $c$ are constants. Because of $R \neq 0$, this type of $f(R)$ theory does not have Schwarzschild's solution.

Where $k^{2}=8 \pi, f(R)$ is a function that only depends on the Rich scalar $R$, and $L_{m}$ is the Lagrangian quantity of matter. For the physical meaning, the function $f(R)$ must satisfy the stability condition(1)There has no ghost particles, $\frac{df}{d R}>0 ;( \mathrm(2))$ There are no tachyon particles, ${d^{2}f}/{d R^{2}}>0$; (3)Stable solution, $\frac{df}{d R}\frac{d^{2} f}{d R^{2}}>R^{[27]}$. In addition, $f(R)$ must also satisfy $\lim _(R \rightarrow \infty)(f(R)-R)/(R)=0$ when there is a valid high-curvature cosmological constant, $\ lim _{R \rightarrow 0} {f(R)-R}/{R}=0$ can return to the early general relativity. Make variation on the action  to get the gravitational field equation
\begin{equation}
G_{\mu \nu} \equiv R_{\mu \nu}-\frac{1}{2} g_{\mu \nu} R=k^{2}\left(\frac{1}{F} T_{\mu \nu}+\frac{1}{k^{2}} \mathcal{T}_{\mu \nu}\right)
\end{equation}where $F=\frac{d f}{d R}, T_{\mu \nu}=\frac{-2}{\sqrt{-g}} \frac{\delta L_{m}}{\delta g^{\mu \nu}}$.
\begin{equation}
\mathcal{T}_{\mu \nu}=\frac{1}{F(R)}\left[\frac{1}{2} g_{\mu \nu}(f-R F)+\nabla_{\mu} \nabla_{v} F-g_{\mu \nu} \square F\right]
\end{equation}
Where $\square=\nabla^{\gamma} \nabla_{\gamma}$.

In the $f(R)$ theory, we consider a general static spherically symmetric black hole, and its geometric form is like this
\begin{equation}
d s^{2}=-W(r) d t^{2}+\frac{d r^{2}}{W(r)}+r^{2} d \Omega^{2}
\end{equation}
This gives the temperature of the black hole,
\begin{equation}
T=\frac{\kappa_{K}}{2 \pi}=\frac{W^{\prime}}{4 \pi}
\end{equation}
where
\begin{equation}
W(r)=1-\frac{2 m}{r}+\beta_{1} r
\end{equation}
Das et al. in \cite{11} have found the form of $\left.\frac{\partial^{2} \mathcal{S}(\beta)}{\partial \beta^{2}}\right|_{\beta=\beta_{0}}=C_{0} T_{H}^{2}$,${S}_{0}= \pi{{r}_{+}}^2$.
\begin{equation}
S=\mathcal{S}_{0}-\alpha (C_{0} T_{H}^{2})
\end{equation}
Here we manually introduce a parameter $\alpha$ to track the corrected term.
The Hawking temperature $\left(T_{H}=\frac{\partial M_{0}}{\partial e^{{S}_{0}}}\right)$ and heat capacity $\left(C_{0}=T_{H} \frac{\partial e^{{S}_{0}}}{\partial T_{H}}\right)$ are calculated as \cite{12}
\begin{equation}
T_{H}=\frac{2 \beta_{1} \pi^{1 / 2} e^{({1 / 2}){S}_{0}}+ \pi}{4 \pi^{3 / 2}  e^{({1 / 2}){S}_{0}}},
\end{equation}
\begin{equation}
C_{0}=-\frac{4 \beta_{1}  \pi^{1 / 2}e^{({3 / 2}){S}_{0}}+2 \pi  e^{{S}_{0}}}{\pi}
\end{equation}

If $r_{+}$denotes the radius of the event horizon, by setting $W(r)=0$, the mass of the black hole, using the relation between entropy $\mathcal{S}_{0}$ and event horizon radius $r_{+},\left(\mathcal{S}_{0}=\pi r_{+}^{2}\right)$, is given by
\begin{equation}
M_{0}\left(\mathcal{S}_{0}, \beta_{1}\right)=\frac{\pi^{1 / 2} \beta_{1} e^{{S}_{0}}+ \pi e^{({1 / 2}){S}_{0}}}{2 \pi^{3 / 2}}
\end{equation}
Ruppeiner geometry is based on fluctuations. When the microstructure of the thermodynamic system under study is unknown, Ruppeiner geometry provides us with a powerful tool for exploring the microstructure of the thermodynamic system. The metric of Ruppeiner geometry is
\begin{equation}
d s^{2}=-\frac{\partial^{2} S(r_{+},\beta_{1})}{\partial X^{\alpha} \partial X^{\beta}} \Delta X^{\alpha} \Delta X^{\beta}
\end{equation}
\begin{equation}
g_{ij}=
\begin{pmatrix}
  2 \pi+\frac{3 \alpha  \beta_{1}  e^{\frac{\pi r_{+} ^{2}}{2}}\left(2 \beta _{1} e^{\frac{\pi r_{+}^{2}}{2}}+\sqrt{\pi}\right)\left(\sqrt{\pi}\left(1+\pi r_{+}^{2}\right)+2 \beta_{1}e^{\frac{\pi r_{+}^{2}}{2}}\left(1+3 \pi r_{+}^{2}\right)\right)}{4 \pi^{3 / 2}} &    \frac{3 \alpha  e^{\frac{\pi r_{+}^{2}}{2}}\left(12 \beta_{1} ^{2} e^{\pi r_{+}^{2}+8\beta_{1}e^{\frac{\pi r_{+}^{2}}{2}} \sqrt{\pi}+\pi}\right) r_{+}}{4 \pi^{3 / 2}} \\
 \frac{3 \alpha e^{\frac{\pi r_{+}^{2}}{2}}\left(12 \beta_{1}^{2} e^{\pi r_{+}^{2}+8 \beta_{1} e^{\frac{\pi r_{+}^{2}}{2}} \sqrt{\pi}+\pi}\right) r_{+}}{4 \pi^{3 / 2}}      &  \frac{3 \mathrm{\alpha } e^{\pi \mathrm{r_{+} }^{2}}\left(2 \mathrm{\beta_{1}  } e^{\frac{\pi \mathrm{r_{+}}^{2}}{2}}+\sqrt{\pi}\right)}{\pi^{5 / 2}}
\end{pmatrix}
\end{equation}
The curvature scalar of thermodynamic geometry is
\begin{equation}
\begin{aligned}
R(S)=&-\left(\left(8 \pi ^ { 5 / 2 } \left(16 \pi^{3}\left(1-7 \pi r_{+}^{2}\right)-96 \beta_{1} e^{\frac{\pi r_{+}^{2}}{2}} \pi^{5 / 2}\left(-1+3 \pi r_{+}^{2}\right)+\right.\right.\right. \\
&\left.3 \alpha\left(2\beta_{1} e^{\frac{\pi r_{+}^{2}}{2}}+\sqrt{\pi}\right)^{2}\left(-24 \beta_{1}^{2} e^{\pi r_{+}^{2}}+\pi\left(-1+\pi^{2} r_{+}^{4}\right)+2 \beta_{1} e^{\frac{\pi r_{+}^{2}}{2} \sqrt{\pi}}\left(-5-4 \pi r_{+}^{2}+3 \pi^{2} r_{+}^{4}\right)\right)\right) / \\
&\left(\left(2 \beta_{1} e^{\frac{\pi r_{+}^{2}}{2}}+\sqrt{\pi}\right)\left(32 \pi^{5 / 2}-3 \alpha\left(2 \beta_{1} e^{\frac{\pi r_{+}^{2}}{2}}+\sqrt{\pi}\right)^{2}\left(\pi^{3 / 2} r_{+}^{2}+2 \beta_{1} e^{\frac{\pi r_{+}^{2}}{2}}\left(-2+3 \pi r_{+}^{2}\right)\right)^{2}\right)\right)
\end{aligned}
\end{equation}
We find the divergence point $2 \beta_{1} e^{\frac{\pi r_{+}^{2}}{2}}=-\sqrt{\pi}$ and
\begin{equation}
32 \pi^{5 / 2}=3 \alpha\left(2 \beta_{1} e^{\frac{\pi r_{+}^{2}}{2}}+\sqrt{\pi}\right)^{2}\left(\pi^{3 / 2} r_{+}^{2}+2 \beta_{1} e^{\frac{\pi r_{+}^{2}}{2}}\left(-2+3 \pi r_{+}^{2}\right)\right)^{2}
\end{equation}

For $R \neq 0$, this type of $f(R)$ theory does not have Schwarzschild's solution.

\section{Thermodynamic of RN black holes in $f(R)$ gravity background}
In this section, we briefly review the main features of the four-dimensional charged AdS black hole corresponding in the $f(R)$ gravity background with a constant Ricci scalar curvature\cite{13}. The action is given by
\begin{equation}
S=\int_{\mathcal{M}} d^{4} x \sqrt{-g}\left[f(R)-F_{\mu \nu} F^{\mu \nu}\right]
\end{equation}

The example we focus on is such a f(R) gravitational background for the reconsidered late acceleration problem\cite{16}
\begin{equation}
f(R)=R-\alpha R^{n}
\end{equation}
The entropy of the RN black hole in this $f(R)$ gravity background is
\begin{equation}
S=\left(1-\alpha n R_{0}^{n-1}\right) \pi r_{+}^{2}=\frac{2-2 n}{2-n} \pi r_{+}^{2}
\end{equation}
$ \alpha$ and $n$ are constants, and $\alpha>0$ and $0<n<1$.
Here, $R$ denotes the Ricci scalar curvature while $f(R)$ is an arbitrary function of $R$. In addition $F_{\mu \nu}$ stands for electromagnetic field tensor given by $F_{\mu \nu}=\partial_{\mu} A_{\nu}-\partial_{\nu} \partial_{\mu}$, where $A_{\mu}$ is the electromagnetic potential. From the action, the equations of motion for gravitational field $g_{\mu \nu}$ and the gauge field $A_{\mu}$ are,
\begin{equation}
\begin{gathered}
R_{\mu \nu}\left[f^{\prime}(R)\right]-\frac{1}{2} g_{\mu \nu}[f(R)]+\left(g_{\mu \nu} \nabla^{2}-\nabla_{\mu} \nabla_{\nu}\right) f^{\prime}(R)=T_{\mu \nu} \\
\partial_{\mu}\left(\sqrt{-g} F^{\mu \nu}\right)=0
\end{gathered}
\end{equation}lytic solution of equation has been determined  when the Ricci scalar curvature is constant $R=R_{0}=$ const, where in this simple case,
\begin{equation}
R_{\mu \nu}\left[f^{\prime}\left(R_{0}\right)\right]-\frac{g_{\mu \nu}}{4} R_{0}\left[f^{\prime}\left(R_{0}\right)\right]=T_{\mu \nu}
\end{equation}
The charged static spherical black hole solution  in $4 \mathrm{~d}$ gravity model takes the form \cite{13}
\begin{equation}
d s^{2}=-N(r) d t^{2}+\frac{d r^{2}}{N(r)}+r^{2}\left(d \theta^{2}+\sin ^{2} \theta d \phi^{2}\right)
\end{equation}
where the metric function $N(r)$ is given by,
\begin{equation}
N(r)=1-\frac{2 m}{r}+\frac{q^{2}}{b r^{2}}
\end{equation}
with $b=f^{\prime}\left(R_{0}\right)$ while the two parameters $m$ and $q$ are proportional to the black hole mass and the charge respectively \cite{14}
\begin{equation}
M=m b, \quad Q=\frac{q}{\sqrt{b}}
\end{equation}
In this background, the electric potential $\Phi$ can be evaluated as
\begin{equation}
\Phi=\frac{\sqrt{b} q}{r_{+}}
\end{equation}
where the black hole event horizon $r_{+}$denotes the largest root of the equation $N\left(r_{+}\right)=0$. At the event horizon, one can also derive the Hawking temperature as well as the entropy of this kind of black hole \cite{13,14},
\begin{equation}
\begin{gathered}
T_{H}=\left.\frac{N^{\prime}\left(r_{+}\right)}{4 \pi}\right|_{r=r_{+}}=\frac{1}{4 \pi r_{+}}\left(1-\frac{q^{2}}{r_{+}^{2} b}\right) \\
S=\pi r_{+}^{2} b
\end{gathered}
\end{equation}
The Hawking temperature $\left(T_{H}=\frac{\partial M_{0}}{\partial {{S}_{0}}}\right)$ and heat capacity $\left(C_{0}=T_{H} \frac{\partial {{S}_{0}}}{\partial T_{H}}\right)$ are calculated.
\begin{equation}
d s^{2}=-\frac{\partial^{2} S(r_{+},n)}{\partial X^{\alpha} \partial X^{\beta}} \Delta X^{\alpha} \Delta X^{\beta}
\end{equation}
\begin{equation}
g_{ij}=
\begin{pmatrix}
 \frac{4(-1+n) \pi}{-2+n} &   -\frac{4 \pi r_{+}}{(-2+n)^{2}} \\
 -\frac{4 \pi r_{+}}{(-2+n)^{2}}     &  \frac{4 \pi r_{+}^{2}}{(-2+n)^{3}}
\end{pmatrix}
\end{equation}
The curvature scalar of thermodynamic geometry is
\begin{equation}
R(S)=\frac{1}{8 \pi r_{+}^{2}}
\end{equation}At that time, when there is a naked singularity, the RN black hole will undergo a phase change.

Das et al. in \cite{11} have found the form of $\left.\frac{\partial^{2} \mathcal{S}(\beta)}{\partial \beta^{2}}\right|_{\beta=\beta_{0}}=C_{0} T_{H}^{2}$,${S}_{0}= \pi{{r}_{+}}^2$.
\begin{equation}
S=\mathcal{S}_{0}-\alpha (C_{0} T_{H}^{2})
\end{equation}
We briefly review the static $f(R)$ black hole solution and its thermodynamics. The generic form of the action (in unit $G=c=\hbar=\kappa=1$ ) is given by
\begin{equation}
I=\frac{1}{2} \int d^{4} x \sqrt{-g} f(R)+S_{\text {mat }}
\end{equation}
among them, $S_{\text {mat }}$ refers to the movement of gravity and the material part of the expression is considered to be
\begin{equation}
f(R)=R-q R^{\beta+1} \frac{\alpha \beta+\alpha+\epsilon}{\beta+1}+q \epsilon R^{\beta+1} \ln \left(\frac{a_{0}^{\beta} R^{\beta}}{c}\right)
\end{equation}
Where $0 \leq \epsilon \leq \frac{e}{4}\left(1+\frac{4}{e} \alpha\right), q=4 a_{0}^{\beta} / c( \beta+1), \alpha \geq 0, \beta \geq 0$ and $a_{0}=l_{p}^{2}, a$ and $c$ are constants.We take the first two parts of the f(R) form.
\begin{equation}
T_{H}=-\frac{2 m\left(1-(\alpha+\alpha \beta+e) q R \theta^{\beta}\right)}{\pi  r_{+}^{3}}
\end{equation}
\begin{equation}
d s^{2}=-\frac{\partial^{2} S(r_{+},q)}{\partial X^{\alpha} \partial X^{\beta}} \Delta X^{\alpha} \Delta X^{\beta}
\end{equation}
\begin{equation}
g_{ij}=
\begin{pmatrix}
 2 \pi-\frac{30 \alpha \left(m-( \alpha+ \alpha \beta +e) m q R \theta^{\beta}\right)^{2}}{\pi r_{+} ^{4}} & -\frac{20 \mathrm{am}\left(\alpha+\mathrm{\alpha }\mathrm{\beta }+\mathrm{e}) \mathrm{q} \mathrm{R} \theta^{\mathrm{\beta}}(m-(\alpha+\mathrm{\alpha }\mathrm{\beta }+\mathrm{e})) \mathrm{q} \mathrm{R} \theta^{\mathrm{\beta}}\right)}{\pi \mathrm{r}_{+} ^{3}}\\
 -\frac{20 \mathrm{am}\left(\alpha+\mathrm{\alpha }\mathrm{\beta }+\mathrm{e}) \mathrm{q} \mathrm{R} \theta^{\mathrm{\beta}}(m-(\alpha+\mathrm{\alpha }\mathrm{\beta }+\mathrm{e})) \mathrm{q} \mathrm{R} \theta^{\mathrm{\beta}}\right)}{\pi \mathrm{r}_{+} ^{3}} & -\frac{10 \alpha \left(-1+(\alpha+\alpha \beta +e) q R \theta^{\beta}\right)^{2}}{\pi r_{+} ^{2}}
\end{pmatrix}
\end{equation}
\begin{equation}
R(S)=\frac{\pi^{3} r_{+} ^{6}}{\left(\pi^{2} r_{+} ^{4}+5 \alpha ^{3}(1+\beta )^{2} m^{2} q^{2} R \theta^{2 \beta}+10 \alpha^{2}(1+\beta) m^{2} q R \theta^{\beta}\left(-1+e q R \theta^{\beta}\right)+5 \alpha  m^{2}\left(-1+e q R \theta^{b}\right)^{2}\right)^{2}}
\end{equation}
We see that there is no divergence of the curvature scalar, and the conclusion is that there is no phase change.R geometry describes the interaction: such as Van der Waals gas, when the molecules are far apart, the interaction is attractive, so the curvature scalar is negative; when the intermolecular distance is short, the interaction is repulsive, and the curvature scalar is positive. Then, for black holes, the gravitational interaction should be very strong.

\section{ Thermodynamic of Kerr-Newman Black Holes in $f(R)$ gravity background}
Since we are looking for a constant curvature $R_{0}$ vacuum solution for the field generated by a large number of charged objects, the appropriate operation (in units of $G=c=\hbar=k_{B}=1$) is:\cite{15}
\begin{equation}
S=\frac{1}{16 \pi} \int \mathrm{d}^{4} x \sqrt{|g|}\left(f(R)-F_{\mu \nu} F^{\mu \nu}\right)
\end{equation}
The metric takes the form:
\begin{equation}
\mathrm{d} s^{2}= \frac{\rho^{2}}{\Delta_{r}} \mathrm{~d} r^{2}+\frac{\rho^{2}}{\Delta_{\theta}} \mathrm{d} \theta^{2}+\frac{\Delta_{\theta} \sin ^{2} \theta}{\rho^{2}}\left[a \frac{\mathrm{d} t}{\Xi}-\left(r^{2}+a^{2}\right) \frac{\mathrm{d} \phi}{\Xi}\right]^{2}
-\frac{\Delta_{r}}{\rho^{2}}\left(\frac{\mathrm{d} t}{\Xi}-a \sin ^{2} \theta \frac{\mathrm{d} \phi}{\Xi}\right)^{2}
\end{equation}
with:
\begin{equation}
\begin{aligned}
\Delta_{r} &:=\left(r^{2}+a^{2}\right)\left(1-\frac{R_{0}}{12} r^{2}\right)-2 M r+\frac{q^{2}}{\left(1+f^{\prime}\left(R_{0}\right)\right)} \\
\rho^{2} &:=r^{2}+a^{2} \cos ^{2} \theta \\
\Delta_{\theta} &:=1+\frac{R_{0}}{12} a^{2} \cos ^{2} \theta \\
\Xi &:=1+\frac{R_{0}}{12} a^{2}
\end{aligned}
\end{equation}
Among them, $M, a$ and $q$ represent the mass, spin and charge parameters respectively.

The entropy  can be expressed as:
\begin{equation}
S=\left(f^{\prime}\left(R_{0}\right)\right) \frac{\mathcal{A}_{H}}{4}
\end{equation}
consequently $1+f^{\prime}\left(R_{0}\right)>0$ is also a mandatory condition to obtain a positive entropy, ${ }^{\mathrm{a}}$ as we supposed above.

\text { Model I: } $f(R)=R+\alpha|R|^{\beta}$
\begin{equation}
R_{0}^{\pm}=\pm\left[\frac{\pm 1}{(\beta-2) \alpha}\right]^{\frac{1}{\beta-1}}
\end{equation}

\text { Model II: } $f(R)=\pm|R|^{\alpha} \exp \left(\frac{\beta}{R}\right)$
\begin{equation}
R_{0}=\frac{\beta}{\alpha-2}
\end{equation}

\text { Model III: } $f(R)=R(\log (\alpha R))^{\beta}$
\begin{equation}
R_{0}=\frac{1}{\alpha} \exp (\beta)
\end{equation}

\text { Model IV: } $f(R)=R-\alpha \frac{\kappa\left(\frac{R}{\alpha}\right)^{n}}{1+\beta\left(\frac{R}{\alpha}\right)^{n}}$
\begin{equation}
R_{0}^{\pm}=-\frac{1-\kappa}{\gamma} \pm \sqrt{\frac{-\kappa(1-\kappa)}{\gamma^{2}}}
\end{equation}

We integrate them into a f(R) gravitational form:
\begin{equation}
f(R)=R-q R^{\beta+1} \frac{\alpha \beta+\alpha+\epsilon}{\beta+1}+q \epsilon R^{\beta+1} \ln \left(\frac{a_{0}^{\beta} R^{\beta}}{c}\right)
\end{equation}
\begin{equation}
d s^{2}=-\frac{\partial^{2} S(r_{+},\beta)}{\partial X^{\alpha} \partial X^{\beta}} \Delta X^{\alpha} \Delta X^{\beta}
\end{equation}
\text { Model I: }
\begin{equation}
g_{ij}=
\begin{pmatrix}
 \frac{(-1+\beta) \pi}{-2+\beta} &   -\frac{ \pi r_{+}}{(-2+\beta)^{2}} \\
  -\frac{ \pi r_{+}}{(-2+\beta)^{2}}     &  \frac{ \pi r_{+}^{2}}{(-2+\beta)^{3}}
\end{pmatrix}
\end{equation}
\begin{equation}
R(S)=\frac{1}{2\pi r_{+}^{2}}
\end{equation}At that time, when there is a naked singularity, the KN black hole will undergo a phase change(Revert to the above RN black hole).

\text { Model II: }
\begin{equation}
g_{ij}=
\begin{pmatrix}
 \pm\left(\frac{\beta}{-2+\alpha}\right)^{-1+\alpha} e^{-2+a} \pi &   \pm\frac{(-1+\alpha)\left(\frac{\beta}{-2+\alpha}\right)^{-1+\alpha} e^{-2+\alpha} \pi r}{\beta} \\
 \pm\frac{(-1+\alpha)\left(\frac{\beta}{-2+\alpha}\right)^{-1+\alpha} e^{-2+\alpha} \pi r}{\beta}     &  \pm\frac{(-2+\alpha)^{2}(-1+\alpha)\left(\frac{\beta}{-2+\alpha}\right)^{\alpha} e^{-2+\alpha} \pi r_{+}^{2}}{2 \beta^{3}}
\end{pmatrix}
\end{equation}
\begin{equation}
R(S)=0.
\end{equation}

\text { Model III: }
\begin{equation}
g_{11}= \frac{1}{2} \pi \log \left[e^{\beta}\right]^{-1+\beta}\left(\beta+\log \left[e^{\beta}\right]\right)
\end{equation}
\begin{equation}
g_{12}=\frac{1}{2} \pi r_{+} \log \left[e^{\beta}\right]^{-2+\beta}\left((-1+\beta)\beta+\log \left[e^{\beta}\right]^{2} \log \left[\log \left[e^{\beta}\right]\right]+\log \left[e^{\beta}\right]\left(1+\beta+\beta \log \left[\log \left[e^{\beta}\right]\right]\right)\right)
\end{equation}
\begin{equation}
g_{21}=\frac{1}{2} \pi r_{+} \log \left[e^{\beta}\right]^{-2+\beta}\left((-1+\beta)\beta+\log \left[e^{\beta}\right]^{2} \log \left[\log \left[e^{\beta}\right]\right]+\log \left[e^{\beta}\right]\left(1+\beta+\beta \log \left[\log \left[e^{\beta}\right]\right]\right)\right)
\end{equation}
\begin{equation}
\begin{aligned}
g_{22}= \frac{1}{4} \pi r_{+}^{2} \log [e^{\beta}]^{-3+\beta}(\beta(2-3 \beta+\beta^{2})+\log [e^{\beta}]^{3} \log[\log [e^{\beta}]]^{2}\\
+\quad \log [e^{\beta}](-2+3 \beta+\beta^{2}+2(-1+\beta)\beta \log[\log [e^{\beta}]])+\\
\log [e^{\beta}]^{2}(2+2(1+\beta) \log[\log[e^{\beta}]]+\beta \log [\log [e^{\beta}]]^{2}))
\end{aligned}
\end{equation}
\begin{equation}
\begin{aligned}
R(S)=(2 \operatorname { L o g } [ e ^ { \beta } ] ^ { 2 - \beta } (\log [e^{\beta}]^{4}(8+3 \log [\log [e^{\beta}]]) \\
+\log [e^{\beta}]^{3}(-21+19 \beta+7(-2+\beta) \log [\log [e^{\beta}]])-(-1+\beta) \beta^{2}(4+5\beta+2\beta \log [\log [e^{\beta}]])+ \\
\quad \beta \log [e^{\beta}](-8+29 \beta-7 \beta^{2}-3(-2+\beta) \beta \log [\log [e^{\beta}]])
+\log [e^{\beta}]^{2}(4-9 \beta+9 \beta^{2}+3 \beta(2+\beta) \log [\log [e^{\beta}]])) / \\
(\pi r_{+} ^ { 2 } ((-1+\beta) \beta^{3}+\log [e^{\beta}]^{4} \log [\log [e^{\beta}]]^{2}+2 \beta \log [e^{\beta}](-2+\beta^{2}+(-1+\beta) \beta \log [\log [e^{\beta}]])+ \\
\quad 2 \log [e^{\beta}]^{3}(-1+(1+\beta) \log [\log [e^{\beta}]]+\beta \log [\log [e^{\beta}]]^{2})\\
+\log [e^{\beta}]^{2}(4-\beta+\beta^{2}+4 \beta^{2} \log [\log [e^{\beta}]]+\beta^{2} \log [\log [e^{\beta}]]^{2}))^{2})
\end{aligned}
\end{equation}
When $\beta=0$ and there is a naked singularity, the KN black hole will undergo a phase change(Revert to the above RN black hole).

\text { Model IV: }
\begin{equation}
R(S)=\infty.
\end{equation}

\section{Summary and Discussion}
There has also been a great development in inferring the possible microstructure of black holes from the macroscopic performance of black hole thermodynamic quantities, such as considering the phase transition of black holes from statistical thermodynamics or inferring the possible phase structures and microscopic interactions of black holes from thermodynamic geometry.Although black holes have the same four laws of thermodynamics as ordinary thermodynamic systems, they are different from ordinary thermodynamic systems, which have many and continue to produce many difficult problems. For ordinary thermodynamic systems, such as solids or gases, the microscopic components are atoms or molecules. Thus, in principle, we can derive its thermodynamic properties from statistical thermodynamics. But for black holes, we do not have the corresponding statistical thermodynamics, nor do we know the microscopic structure of black holes, and what is worse, even whether black holes have microscopic structures is unclear. Perhaps the black hole itself is a giant elementary particle.

In the context of $f(R)$ modified gravitational theory, we study the Kerr-Newman black hole solution. We study the non-zero constant scalar curvature solution and discuss the metric tensor that satisfies the modified field equation. We have studied the thermodynamics of BH and its local and global stability without the cosmological constant. We analyzed these characteristics in various $f(R)$ models. We commented on the main differences in general relativity and demonstrated the rich thermodynamic phenomenology that characterizes the framework.

In this article, we consider a conventional black hole under f(R) gravity. Due to the small fluctuations around the equilibrium state, we have summarized the expression of the modified thermodynamic entropy of this black hole. At the same time, we also studied the geometric thermodynamics (GTD) of black holes, and studied the adaptability of the curvature scalar of the geometric thermodynamics method to the phase transition point of the black hole. In addition, we also studied the influence of the modified parameters on the thermodynamic behavior of black holes.

{\bf Acknowledgements:}\\
This work is partially supported by  National Natural Science Foundation of China(No. 11873025).

\end{document}